\documentclass{PoS}

\usepackage{epsfig}
\usepackage{dsfont}

\PoS{PoS(LAT2005)127}

\title{Low-lying spectrum for lattice Dirac operators with twisted mass}

\ShortTitle{Spectrum for twisted mass fermions}

\author{\speaker{Christof Gattringer}\\

	Institut f\"ur Physik, FB Theoretische Physik\\
        Karl-Franzens-Universit\"at Graz\\
	A-8010 Graz, Austria.\\

        E-mail: \email{christof.gattringer@uni-graz.at}\\
       }

\author{Stefan Solbrig\\

	Institut f\"ur Theoretische Physik\\
        Universit\"at Regensburg\\
	D-93040 Regensburg, Germany.\\

        E-mail: \email{stefan.solbrig@physik.uni-r.de}\\
       }

\abstract{
We analyze the low-lying spectrum and eigenmodes of lattice Dirac operators
with a twisted mass term. The twist term expels the eigenvalues from a strip
in the complex plane and all eigenmodes obtain a non-vanishing matrix element
with $\gamma_5$. For a twisted Ginsparg-Wilson operator the spectrum is
located on two arcs in the complex plane. Modes due to non-trivial topological
charge of the underlying gauge field have their eigenvalues at the edges of
these arcs and obey a remnant index theorem. For configurations in the
confined phase we find that the twist mainly affects the zero modes, while the
bulk of the spectrum is essentially unchanged.
}

\FullConference{XXIIIrd International Symposium on Lattice Field Theory\\
		 25-30 July 2005\\
		 Trinity College, Dublin, Ireland}

\begin{document}

In the last few years lattice QCD formulations with a twisted mass term were
studied in great detail \cite{twist1}-\cite{GaSo05} and the latest results
were reviewed at this conference \cite{twistrev}. The interest of the lattice
community in twisted mass QCD is based on the fact that the twisted mass term 
protects the Dirac operator from small eigenvalues, thus avoiding exceptional 
configurations. This allows one to work at small quark masses and the results
can be extrapolated to vanishing twist.

However, the fact that no small eigenvalues appear implies that also
zero modes of the Dirac operator are excluded. Zero modes of the Dirac
operator in turn are, via the index theorem \cite{index}, 
related to infrared structures of the gauge field
carrying topological charge. This connection already
raises the question we address in this contribution: How does topological
charge of the gauge field manifest itself in the eigensystem of the Dirac
operator with a twisted mass term?

We stress that this question is not only of academic interest, since
topological objects are believed to play an important role in the QCD
vacuum. Thus it should be understood how topological objects 
interact with the fermionic degrees of freedom in the presence of a 
twisted mass term.

For two mass-degenerate flavors of fer\-mi\-ons, $u$ and $d$, 
the twisted mass Dirac operator ${\cal D}$ has the form 
$$
{\cal D} \; = \; \big( D_0 \, + \, m \big) \mathds{1}_2 \; + \; 
i \,\mu \, \gamma_5 \, \tau_3 \; \;\; \; , \; \; \; \; \; \; 
\tau_3 \, = \, \mbox{diag}\,(1,-1) \, ,
$$
where $\mathds{1}_2$ and $\tau_3$ act in flavor space. For our analysis it 
is convenient to analyze the Dirac operators for $u$ and $d$ separately 
by setting $D_u = D(\mu)$ and $D_d = D(-\mu)$ with
\begin{equation}
D(\mu) \; = \; D_0 \, + \, m \, + \, i \, \mu \, \gamma_5 \; .
\label{dmudef}
\end{equation} 
The Dirac operator $D_0$ is the 1-flavor Dirac operator, discretized on the
lattice without doublers. A common choice is Wilson's Dirac operator, but here
we will also consider (for theoretical purposes) a Dirac operator that is 
chiral, i.e., obeys the Ginsparg-Wilson relation \cite{giwi}. 

A property which we assume for the operator $D_0$, and which is obeyed by 
Wilson's Dirac operator and also the overlap operator, is 
$\gamma_5$-hermiticity,
\begin{equation}
\gamma_5 \, D_0 \, \gamma_5 \; = \; D_0^\dagger \; .
\label{gamma5herm}
\end{equation}
For the twisted operator $D(\mu)$ this property implies the 
generalized $\gamma_5$-hermiticity relation
\cite{generg5herm}
\begin{equation}
\gamma_5 \, D(\mu) \, \gamma_5 \; = \; D(- \mu)^\dagger \; .
\label{gamma5hermgen}
\end{equation}
The generalized $\gamma_5$-hermiticity relation can be explored to derive
results for the eigensystem of the Dirac operator. We denote an eigenvector 
of $D(\mu)$ with eigenvalue $\lambda$ by $\psi_\lambda$ and find for its
matrix element with $\gamma_5$:

\begin{eqnarray}
&& \hspace{-7mm}
\lambda (\psi_\lambda, \gamma_5 \psi_\lambda) = 
(\psi_\lambda, \gamma_5 D(\mu) \psi_\lambda) = 
(\psi_\lambda, D(-\mu)^\dagger \gamma_5 \psi_\lambda) = 
(D(-\mu) \psi_\lambda, \gamma_5 \psi_\lambda) = 
\nonumber 
\\
&& \hspace{-7mm}
([D(\mu)\! - \! i 2 \mu \gamma_5 ] \psi_\lambda, \gamma_5 \psi_\lambda) =
\lambda^* (\psi_\lambda, \gamma_5 \psi_\lambda) \! + \! i 2 \mu 
(\psi_\lambda, \psi_\lambda) = 
\lambda^* (\psi_\lambda, \gamma_5 \psi_\lambda) \! + \! i 2 \mu .
\end{eqnarray}
We have used (\ref{gamma5hermgen}) and the fact
that the eigenvectors are normalized to 1. From the last equation
follows 
\begin{equation}
(\psi_\lambda, \gamma_5 \psi_\lambda) \; = \; \frac{\mu}{\mbox{Im} \,
\lambda} \; .
\label{matrixelement}
\end{equation}
This formula implies that all eigenvectors have non-vanishing chirality which
decreases monotonically with $\mbox{Im} \, \lambda$. This is very different
from the case without twist term, where only eigenmodes with real eigenvalue
have a non-vanishing matrix element with $\gamma_5$. These modes correspond to 
zero modes in the continuum and are of topological nature.

Using the fact that $\gamma_5$ is bounded, i.e.,
$|(\psi_\lambda, \gamma_5 \psi_\lambda)| \leq 1$, we find ($\mu \geq 0$)
\begin{equation}
\mid \mbox{Im} \, \lambda \mid \; \geq \; \mu \; .
\label{imaginarybound}
\end{equation}
Thus all eigenvalues have an imaginary part larger or equal the twist
parameter $\mu$ (positive in our convention). 
The twist term expels the spectrum from a strip of width 
$2\mu$ along the real axis and thus the smallest eigenvalues have a modulus 
with a size of at least $\mu$.

\begin{figure}[t]
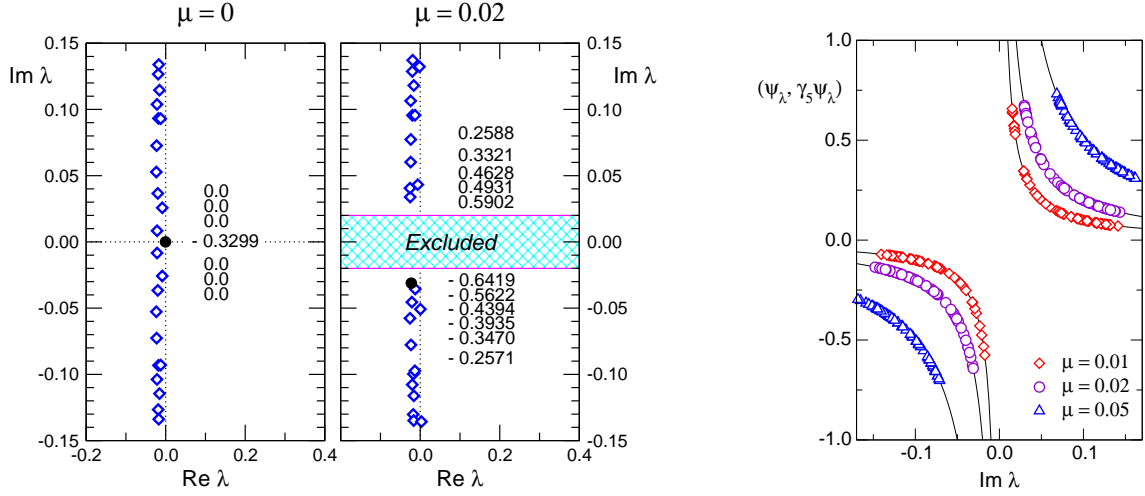

\begin{center}
\includegraphics[height=6.5cm,clip]{sp_b845_16x16_c6_DWilson.eps}
\hfill
\includegraphics[height=6.1cm,clip]{g5_vs_imlambda_16x16_b845_Wi.eps}
\caption{Left-hand side plot: Spectra of the Dirac operator in the 
  complex plane
  for $\mu = 0.0$ and $\mu = 0.02$. The numbers next to the symbols
  for the eigenvalues are the matrix elements of the corresponding
  eigenvectors with $\gamma_5$.  Right-hand-side plot: $\gamma_5$ 
  matrix elements of the eigenvectors $\psi_\lambda$ as a function of 
  Im$\lambda$. The full curves are the hyperbolas of Eq.\ (5). }
\end{center}
\end{figure}

The basic properties of the spectrum and the eigenmodes are illustrated in
Figure 1 which was generated using the twisted Wilson Dirac operator on a
quenched gauge field configuration of topological charge $Q = +1$ on a 
lattice of size $16^4$ at a lattice spacing of $a = 0.094$ fm as 
determined from the Sommer parameter \cite{scale}.

In the left-hand side plot we compare the spectra in the complex plane 
for vanishing twist and at a twist of $\mu = 0.02$ in lattice units. 
As the twist term is turned on, the eigenvalues are shifted away from the real
axis. The topological eigenvalue (filled symbol) is shifted in the negative
imaginary direction. It is obvious that the spectrum is no longer symmetric
with respect to reflection at the real axis. We remark that the fluctuations
of the eigenvalues in horizontal direction are not reduced by the twist term.
The numbers next to the symbols
representing the eigenvalues $\lambda$ are the values for the matrix elements
$(\psi_\lambda, \gamma_5 \psi_\lambda)$ of the corresponding eigenvectors
$\psi_\lambda$. 

In the right-hand side plot of Figure 1 we show these
matrix elements as a function of Im$\lambda$ for three different values of 
$\mu$. The symbols represent the numerical data, the full curves show the
hyperbolas of Eq.\ (5).

As discussed, the focus of this paper is on the low lying spectrum of the
twisted mass lattice Dirac operator and the role of the index theorem. For the
index theorem it 
is important to use a setting where chiral symmetry is properly implemented on
the lattice and to consider the role of the twisted mass term in this case.
Thus we require now
the Dirac operator $D_0$ (before the twist term is turned on) 
to obey the Ginsparg-Wilson relation,
\begin{equation}
\gamma_5 D_0 \, + \, D_0 \gamma_5 \; = \; a D_0 \gamma_5 D_0 \, ,
\label{giwi}
\end{equation}
which governs chiral symmetry on the lattice. In this case $D_0$
can have exact zero modes $\psi_0^\pm$ which are chiral, i.e., 
$\gamma_5 \psi_0^\pm = \pm \psi_0^\pm$. They are related to the topological
charge $Q$ of the gauge field via the index theorem,
\begin{equation}
Q \; = \; n_- \, - \, n_+ \; ,
\label{indextheoremlat}
\end{equation}
where $n_+$ and $n_-$ are the numbers of right- and left-handed zero modes
\cite{giwiindex}.

Since the zero modes $\psi_0^\pm$ of $D_0$ are chiral, they are also eigenmodes
of $D(\mu)$ with eigenvalue $\pm i \mu$, with a plus sign for the right-handed
mode and a minus sign for a left-handed zero mode. Thus in the presence of the
twisted mass term we can identify a remnant of the index theorem which still
has the form of Eq.\ (\ref{indextheoremlat}), but $n_-$ and $n_+$ are now
interpreted as the number of eigenvalues $+ i \mu$ and $- i \mu$. 

\begin{figure}[t]
\begin{center}
\includegraphics[width=13cm,clip]{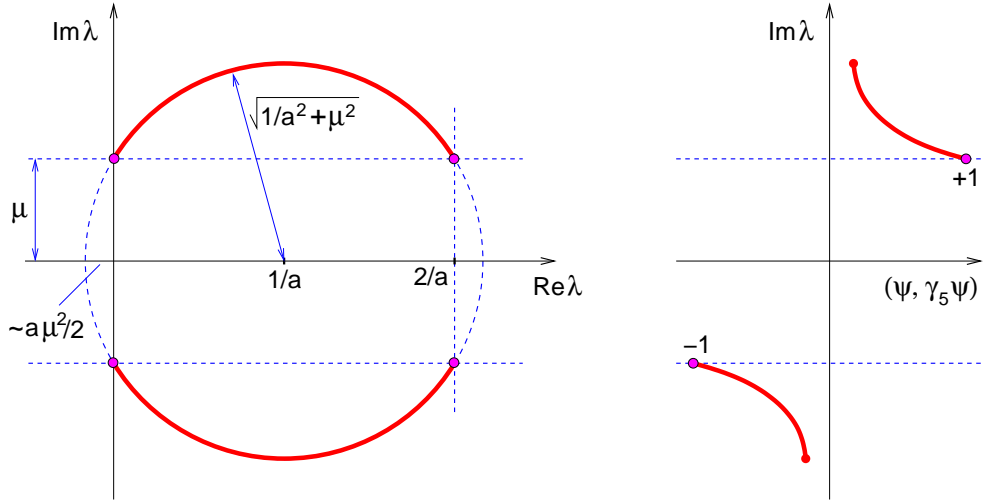}
\caption{Schematic representation of our results for spectrum (left-hand side 
plot) and the matrix element with $\gamma_5$ (right-hand side plot) for a 
chiral Dirac operator with twisted mass term.}
\end{center}
\end{figure}

From (\ref{dmudef}) one finds $D_0 = D(\mu) - i \mu \gamma_5$ and inserting 
this expression into (\ref{giwi}) one obtains after a few lines of
algebra (use (\ref{gamma5hermgen}) and $D(-\mu) = D(\mu) - i 2 \mu \gamma_5$), 
\begin{equation}
D(\mu) \, + \, D(\mu)^\dagger \; = \; a D(\mu) D(\mu)^\dagger \, - \, a \mu^2
\; .
\end{equation}
Sandwiching this equation between eigenvectors $\psi_\lambda$ gives rise to
the equation 
\begin{equation}
\lambda \, + \, \lambda^* \; = \; a \lambda \lambda^* \, - \, a \mu^2 \; ,
\label{circle}
\end{equation}
for the corresponding eigenvalue $\lambda$.
Setting $\lambda = x + i y$ one finds that this is the equation
for a circle in the complex plane. This ``stretched Ginsparg-Wilson circle''
has center $1/a$ and radius $\sqrt{1/a^2 + \mu^2}$. Since $|Im \lambda|
\ge \mu$ we find that the spectrum of $D(\mu)$ is confined to the two arcs
shown in the left-hand side plot of Fig.~2. The topological eigenmodes with
eigenvalues $\pm i \mu$ fall one the edges of the two arcs (on the other edges
their doubler partners are found). The matrix elements of the eigenvectors
with $\gamma_5$ still obey Eq.~(\ref{matrixelement}). The right-hand side
plot of Fig.~2 illustrates the results for the $\gamma_5$-matrix element.

\begin{figure}[t]
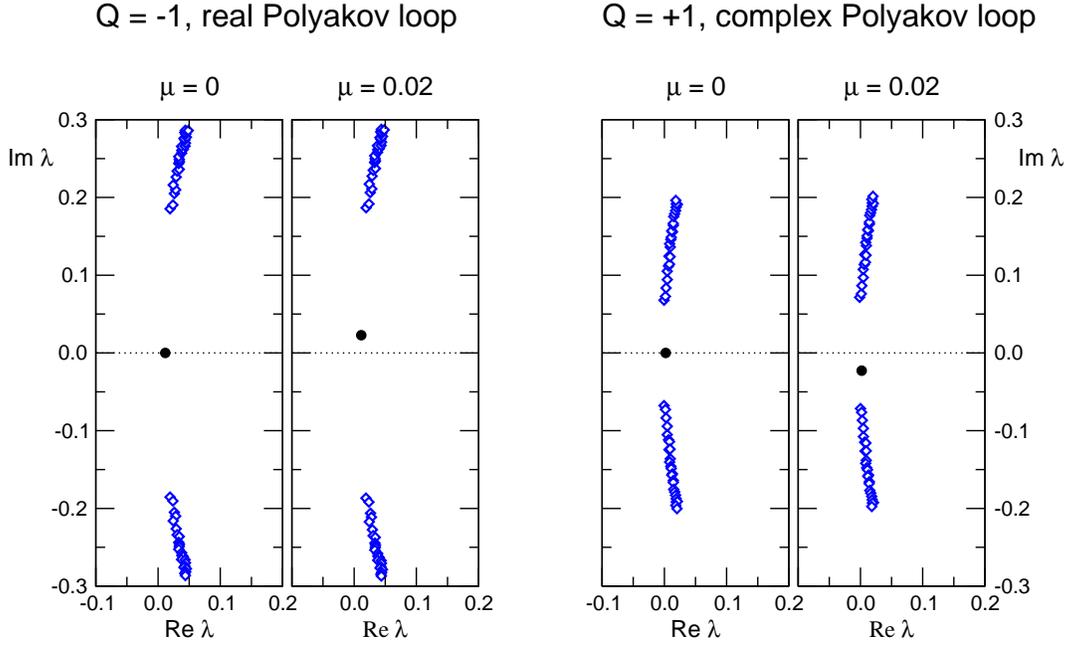

\begin{center}
\includegraphics[height=8.5cm,clip]{sp_fintemp_realP.eps}
\hspace{1cm}
\includegraphics[height=8.5cm,clip]{sp_fintemp_complexP.eps}
\caption{Spectrum of the Dirac operator in the deconfined, chirally symmetric,
phase. We compare the untwisted case to the situation at $\mu = 0.02$. The
left-hand side plot is for a configuration with $Q = -1$ 
and real Polyakov loop,
while the right-hand side plot is for $Q = 1$ and complex Polyakov loop.}
\end{center}
\end{figure}

Our discussion of a Ginsparg-Wilson operator with a twisted mass term shows
that the zero modes of a chiral Dirac operator without twist, $D_0$, 
are protected by chiral symmetry and remain unchanged as the twist term 
is turned on (only the eigenvalue changes). Instead, when using the Wilson
Dirac operator for $D_0$, this protection is only approximate. In 
\cite{GaSo05} characteristic properties of the topological zero mode (edge
mode), such as its localization and $\gamma_5$ matrix element, 
were studied for different values of $\mu$. While for the twisted Wilson 
operator these properties changed considerably, for the approximately 
chiral CI operator \cite{CIref} they remained essentially unchanged.

Another question that was addressed in \cite{GaSo05} is the role of the
twisted mass term in the deconfined (chirally symmetric) phase. In the
chirally symmetric phase the spectrum of the Dirac operator has a gap near the
origin (up to isolated zero modes) which implies, via the Banks-Casher
relation \cite{baca}, a vanishing chiral condensate. Turning on the 
twisted mass term mainly affects the zero modes. As seen in Fig.~3, they are
shifted up or down, depending on their chirality. The size of the spectral
gap remains essentially unchanged.

\end{document}